\documentclass[a4paper,12pt]{article}
\usepackage[dvipsnames]{xcolor}
\usepackage[utf8]{inputenc} 
\usepackage[hidelinks]{hyperref}
\usepackage{float}
\usepackage{amssymb,amsmath,amsthm}
\usepackage{amsfonts}

\usepackage[draft,colorinlistoftodos, textwidth=3cm, textsize=scriptsize]{todonotes}

\pdfoutput=1
\usepackage{graphicx}
\usepackage{graphicx, float, tabularx, booktabs}
\usepackage{color}
\usepackage{enumerate}
\usepackage[american]{babel}
\usepackage{stackrel}

\usepackage[numbers, compress]{natbib}
\usepackage{natbib}
\usepackage{slashed}
\usepackage{amsfonts}
\usepackage{epsfig}
\usepackage{graphicx, tabularx}

\newcommand{\MF}{M_\mathrm{F}}
\newcommand{\MP}{M_\mathrm{P}}
\newcommand{\LP}{L_\mathrm{P}}
\newcommand{\be}{\begin{equation}}
\newcommand{\ee}{\end{equation}}
\newcommand{\ba}{\begin{eqnarray}}
\newcommand{\ea}{\end{eqnarray}}

\vspace{-0.1cm}
\renewcommand{\thefootnote}{\fnsymbol{footnote}}

\newcommand{\gsim}{\mathrel{\hbox{\rlap{\lower.55ex \hbox {$\sim$}}
                   \kern-.3em \raise.4ex \hbox{$>$}}}}
\newcommand{\lsim}{\mathrel{\hbox{\rlap{\lower.55ex \hbox {$\sim$}}
                   \kern-.3em \raise.4ex \hbox{$<$}}}}

 \usepackage{graphicx}

\usepackage{amsmath}

\newcommand{\bw}{\begin{widetext}}
\newcommand{\ew}{\end{widetext}}

\newcommand{\der}{\ensuremath{ d }}

\def\ber{\begin{eqnarray}}
\def\eer{\end{eqnarray}}

\begin{document}

\begin{center}
{\Large \textbf{Quantum gravity and the zero point length
}
}
\end{center}

\vspace{-0.1cm}

\begin{center}
Piero Nicolini$^{a,b}$\footnote{%
E-mail: \texttt{nicolini@fias.uni-frankfurt.de} }

\vspace{.6truecm}

\emph{\small  $^a$Frankfurt Institute for Advanced Studies (FIAS),\\[-0.5ex]   Frankfurt am Main, Germany}\\[1ex]

\emph{\small  $^b$Institut f\"ur Theoretische Physik,\\[-0.5ex]  
Goethe-Universit\"at, Frankfurt am Main, Germany}\\[1ex]

\end{center}
\begin{abstract}
\noindent{\small  
In this paper, we present an overview of some of the existing issues of the research in quantum gravity. We also introduce the basic ideas that led  Padmanabhan to consider a duality property in path integrals. Such a duality is consistent  with the T-duality in string theory. More importantly, the path integral duality discloses a universal feature of any quantum  geometry, namely the existence of a zero point length $L_0$. We also comment about recent developments aiming to expose effects of the zero point length in strong electrodynamics and black holes.   There are reasons to believe that the main characters of the phenomenology of  quantum gravity may be described by means of a single parameter like $L_0$. 
}
\end{abstract}

\renewcommand{\thefootnote}{\arabic{footnote}} \setcounter{footnote}{0}

\section{Introduction}

\begin{flushright}
\begin{scriptsize}

\textit{Any approach towards the quantum structure of spacetime is\\   useful if and only if it has
something sensible to say about:\\
i) the  singularity in cosmology;\\
ii) the singularity in black hole spacetimes;\\
iii) the cosmological constant problem.\\}
-- T. Padmanabhan, Frankfurt, 2013
\end{scriptsize}
\end{flushright}

If we had to define quantum gravity in plain words, we could say it is an umbrella term for an array of theoretical paradigms aiming to describe gravity at quantum level. From a broader perspective,  quantum gravity is, however, more than this. It is likely a mandatory step towards the full understanding of fundamental interactions governing the Universe from sub-atomic to cosmological scales. Therefore, we should better say that quantum gravity is an important part of the \textit{theory of everything} or, more simply, it actually coincides with it.

Despite being the Everest of physics,  despite the tremendous theoretical efforts, e.g. the formulation of Superstring Theory,  Loop Quantum Gravity and other candidate theories (for an overview see \cite{Kie12}), despite the investments to support dedicated research programs and institutes,  research in quantum gravity has not had much progress during the last fifty years. Such a disappointing development is connected to three main obstacles.  

Contrary to quantum field theory, quantum gravity formulations  are, in general, mathematical intractable or, in the best case, hard to work out. This feature has a negative consequence, that is actually the second limitation of quantum gravity: a certain scarcity of phenomenological predictions. As of today, the Hawking radiation can be regarded as the most striking indication emerging from the combination of gravity and quantum mechanics \cite{Haw74,Haw75}. Hawking's result  is, however, dated back to 1974 and stems from quantum field theory in curved space, rather than from the aforementioned candidate theories of quantum gravity. There is also a third problem that is by far the most alarming. As of today, no quantum gravity signal has yet been detected.  For this reason,  competing candidate theories can coexist without any possibility of  experimentally discerning the correct one\footnote{For further considerations about quantum gravity, I recommend the reading of the review of H. Nicolai \cite{Nicolai13}}.

Such a situation is ever more dramatic if one thinks that quantum gravity is expected to pave the way towards the clarification of an array of unsolved problems in physics\footnote{Padmanabhan offered illuminating remarks about this point during his keynote at the 1st Karl Schwarzschild Meeting \cite{Pad16} -- see the quotation at the beginning of the paper.}. These include the hierarchy problem, the trans-Planckian problem, the curvature singularity problem, the cosmological constant problem, the black hole information paradox and the problem of the nature of dark matter and dark energy. All the above problems  result from the non-fundamental nature of two basic pillars of theoretical physics, namely general relativity and the Standard Model of particle physics. In practice, such pillars are just effective formulations of something we currently do not master \cite{tGR18}. 

Quantum gravity has a special role in physics also for an additional reason: It is not just like a ``normal'' nonrenormalizable theory. By normal, I mean a theory that can be described in an effective way  below a certain energy scale. Above such a scale, the effective theory breaks down and one necessarily needs the full theory. 
It has been noted by many authors, e.g. Amati, Ciafaloni \& Veneziano \cite{ACV87,ACV88,ACV89}, 't Hooft \cite{tHo90}, Maggiore \cite{Mag93},  Aurilia  \& Spallucci \cite{AuS02,AuS13}, Carr \cite{Carr16,CMN15,CMMN20} and more recently Dvali \& Gomez \cite{DvG10,DFG11}, that gravity does have a  characteristic scale, but does not follow the above scheme for energies above or below it. Indeed, above its characteristic  scale, gravity becomes classical! One finds that general relativity and quantum field theory in curved space can efficiently  describe all phenomena at such energies.

 Such a conclusion is absurd at first sight, but it can be explained in simple terms as follows. In the absence of 
 extradimensions and particular spacetime topology, the characteristic scale of gravity is the Planck scale, $\MP\equiv \sqrt{1 /G}\sim 10^{19}$ GeV. At such an energy scale, gravity is no longer weak and particles undergo a gravitational collapse into
  microscopic black holes. Even if the details of such a collapse are not fully understood, its occurrence implies the existence 
  of a minimal length for particle physics, namely $\LP=1/\MP$. Such a minimal length is both the Compton 
  wavelength of the collapsing particle, $\sim 1/M$ and the gravitational radius of the forming black hole, $\sim GM$. The two coincides only if $M=\MP$, i.e. the mass value at which the   particle--black hole system is realized. A further increase of energy, $M>\MP$, will not lead to a smaller particle but rather   to a larger black hole, whose size is governed by the gravitational radius only. In practice gravity turns to be \textit{self-complete}: Whatever is the energy deployed, gravity makes length scales below $\LP$ inaccessible. 
  
  \begin{figure}[h]
\begin{center}
\includegraphics[width=0.7\textwidth]{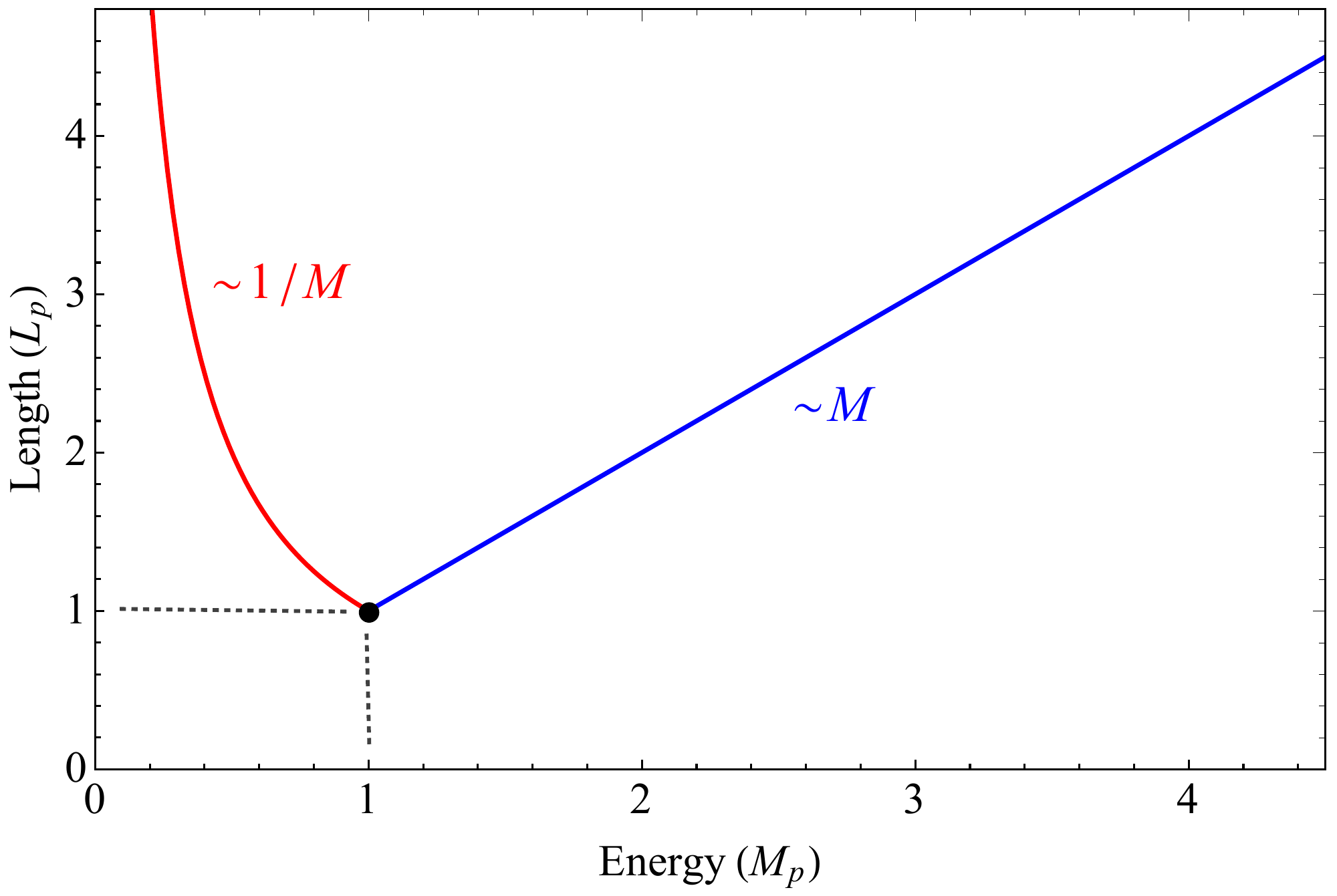}
\end{center}
\caption{\small
Gravity self-completeness. There exist two phases of matter, i.e. particles and black holes. At the Planck scale there is the transition between the two.}
\label{fig:fig1}
\end{figure}

If on the one hand, gravity self-completeness downplays the problem of curvature singularities, on the other hand, one is left with the even bigger problem of the regime of quantum gravity. In the length vs. energy diagram in Fig. \ref{fig:fig1}, quantum gravity corresponds just to a point (black spot), i.e.,  the Planck scale $(\MP,\LP)$. This is the visual representation of the  ``non-normal'' character of gravity, as mentioned above. In practice, the hierarchy problem, i.e. the extremely high value of the Planck mass, is worsened by the fact that the quantum regime of gravity around $\MP$ is quite narrow.   For this reason, there is a general  difficulty in formulating effective field theories able to faithfully reproduce quantum gravity phenomenology at scales other than the Planck scale \cite{Vaf05}. 

Despite the above list of long standing problems,  there exist positive aspects of the current research on quantum gravity. Actually there is something we know rather well.  I am referring to a basic result that has been derived by Padmanabhan on the ground of general quantum gravity arguments \cite{Pad97}. 
  In the following section, I will briefly summarize the kernel of Padmanabhan's reasoning for the derivation of such a result. In Sec. \ref{sec:pheno}, I will focus on some of the applications that could lead to the exposure of distinctive effects of quantum gravity. In Sec. \ref{sec:concl}, I will offer some concluding remarks. Finally, in Sec. \ref{sec:IandPaddy}, I will recall something in relation to my interaction with Padmanabhan.

\section{Zero point length and path integral duality}

We start by considering matter propagating on a background spacetime along the lines of what presented in \cite{Pad97,Pad20}.

The amplitude for a relativistic particle is proportional to $\exp(-A)$, where $A$ is the action\footnote{For simplicity we have assumed a Euclidean spacetime.}.
We can write $A=m\sigma=\sigma/\lambda$, where $\sigma$ is the path, $m$ is the particle mass and $\lambda$ is the Compton wavelength.  The dynamics of the related quantum field is described by the propagator, whose path integral representation reads:
\begin{equation}
G(x,y; m^2)=\sum_{\mathrm{paths}\ \sigma}\exp\left[-\sigma(x,y)/\lambda\right].
\label{eq:pathint}
\end{equation}
Here $\sigma(x,y)$ is a path connecting the points $x$ and $y$ of the spacetime lattice. The sum is carried over the discrete set of all paths that connect the above two points. In the continuous limit, one obtains the conventional Schwinger representation of the propagator. It is interesting to notice that paths longer than $\lambda$ are suppressed, i.e., they marginally affect the quantum dynamics.

At this point, Padmanabhan's line of reasoning is the following. The spacetime manifold  should be considered just as a large distance limit of the actual quantum spacetime. As a result, the transition from the discrete to the continuum should have a memory of the fact that a quantum manifold fluctuates. Similarly to the spectrum of Hamiltonian operators, the length of paths on such a manifold should never vanish. Rather, such paths must  display a \textit{zero point length}, $L_0$, due to the intrinsic quantum uncertainty.  This is translated in the path integral by a term that suppresses paths shorter  than $L_0$. In other words,  the effects of quantum gravity can be taken into account by the substitution    
\begin{equation}
\sigma\longrightarrow \sigma + \frac{L_0^2}{\sigma} 
\label{eq:zeropoint}
\end{equation}
in \eqref{eq:pathint}. The above modification is universal and purely geometrical. It is simply the response of the spacetime, irrespective of the kind of matter propagating on it. It does not depend on the spacetime curvature either. Given the minuscule size of $L_0$, the spacetime will actually appear locally flat in the majority of cases, i.e., $L_0\ll (K)^{-1/2}$, being $K$ the Kretschmann scalar. 

The combination of \eqref{eq:pathint} and \eqref{eq:zeropoint} offers further insights about the nature of quantum gravity. There exists a new symmetry between the world ``above $L_0$'' (i.e. $\sigma > L_0$) and that ``below $L_0$'' (i.e. $\sigma < L_0$). Such a symmetry has been observed in the context of string theory and has been termed target space duality, or simply T-duality \cite{GPR94,Lus10}. It implies the equivalence of theories defined on compact spaces of size $R$ and those defined on compact spaces of size proportional to $1/R$.

The propagator emerging from \eqref{eq:pathint} and \eqref{eq:zeropoint} does not only share a symmetry with string theory. It is more than a simply analogy. The propagator can be actually derived in an alternative way in the context of bosonic closed string in $(3+1)+1$-dimensions. Padmanabhan and Spallucci, in collaboration with Smailagic \cite{SSP03} and Fontanini \cite{SpF05,FSP06}  showed this is actually possible and were able to identify the zero point length with the string parameter $\alpha^\prime$, namely
\begin{equation}
L_0=2\pi\sqrt{\alpha^\prime}.
\end{equation}  

The contact with string theory is very intriguing because it offers a first principle justification for the substitution in \eqref{eq:zeropoint}. String theory, however, does not set the scale, since $\alpha^\prime$ is a free parameter. To circumvent such a problem, Padmanabhan has proposed also a third derivation of result in  \eqref{eq:zeropoint}. 

Instead of relying on arguments stemming from properties of the spacetime, we can alternatively consider an intrinsic feature of the Compton wavelength. In simple terms, such a wavelength indicates the minimal size, $\lambda$, where a single particle of mass $m$ can be localized. This is a crucial scale for creation and annihilation processes in quantum field theory. To observe lengths smaller than $\lambda$, one should illuminate the particle with photons of energy $E$, such that $1/E<\lambda$. This implies $E>m$ and consequent creation of additional particles. In practice, a single particle of mass $m$ cannot fit into box of edge $1/E$, with $E>m$. 

At this point Padmanabhan has made a simple but powerful consideration. The relation $\lambda =1/m$ would suggest that the heavier is the particle, the smaller is the Compton wavelength. This is correct but it can be valid for a certain mass regime only. At a certain mass scale, gravity necessarily becomes important for the particle itself and can even lead to matter collapse. To avoid such a scenario, one has to make sure that paths smaller than the gravitational radius of the particle, $r_\mathrm{g}$, do not contribute to the path integral \eqref{eq:pathint}. Accordingly, the most natural way to amend the standard path integral is:
\begin{equation}
\sum_{\mathrm{paths}\ \sigma}\exp\left[-\sigma(x,y)/\lambda\right]\longrightarrow \sum_{\mathrm{paths}\ \sigma}\exp\left[-\sigma(x,y)/\lambda\right]\exp\left[-r_\mathrm{g}/\sigma(x,y)\right].
\end{equation}
The additional term implies the same modification emerging from the introduction of the zero point length \eqref{eq:zeropoint} in the propagator \eqref{eq:pathint}, provided $L_0^2=G$, being $r_\mathrm{g}\sim Gm$. Therefore, we can conclude that $L_0$ has to be of the order of the Planck length.

It is important to remark that the above derivation is based on the tacit assumption that the Principle of Equivalence remains valid at the Planck scale. Indeed, $\lambda$ depends on the particle inertial mass. Conversely, $r_\mathrm{g}$ depends on the particle gravitational mass. Only if the two masses coincide, $L_0$ universally assumes Planckian values, i.e., irrespective of the properties of the particle under consideration.

\section{Phenomenology of quantum gravity}
\label{sec:pheno}

The prediction of effects emerging from quantum gravity has been one  the hottest topics in physics for the last twenty years (see some existing reviews \cite{Ame13,SNB12,Hos13}).

The general strategy to expose putative quantum gravity effects is to consider ultra high energy events. This is for instance the case of some  cosmic rays whose  energy can exceed $10^{11}$ GeV. Despite very large detectors have been put in operation during the last years, fluxes of cosmic rays at such energy regimes remain too meager \cite{TelescopeArray:2014tsd}. To overcome such a limitation, there are two possible alternatives. 
Instead of a single high energy event, small quantum gravity effects can be accumulated by weakly propagating probes traveling over cosmological distances. For instance, neutrino oscillation wavelength can be elongated by the interaction with the quantum spacetime \cite{SBN11}. As a second alternative, one can modify the theoretical scenario and postulate a mechanism
to solve the hierarchy problem. In practice, one aims to lower the fundamental scale, $\MF$, from a value close to the Planck scale to another value  in the vicinity of the electroweak scale, e.g., $\MF\sim 1$ TeV. Spacetime models with additional spatial dimensions (e.g. large extra dimensions \cite{AAD+98,ADD98,ADD99,BaF99}, brane-world \cite{Gog98a,Gog98b,Gog99,RaS99a,RaS99b}, universal extradimensions \cite{ACD01})  are among the most popular conjectures belonging to such a second alternative.

In the following paragraphs, we will focus on the phenomenological repercussions of Padmanabhan's propagator duality. In doing so, we will benefit from two good properties. First,  the propagator \eqref{eq:pathint} can analytically be computed. In the continuous limit in flat space it reads 
\begin{equation}
G(x,y; m^2)=\int \frac{d^D p}{(2\pi)^D}e^{-ip\cdot(x-y)} G(p),
\end{equation}
with
\begin{equation}
G(p)= -\frac{L_0}{\sqrt{p^2+m^2}}\, K_1 \left(L_0 \sqrt{p^2+m^2}\right),
\label{eq:paddyprop}
\end{equation}
where $K_1(x)$ is a modified Bessel function of the second kind. For small arguments $x\ll1$, one has $K_1(x)\sim x^{-1}$. This leads to  the conventional quantum field theory result for $G(p)$. For large arguments $x\gg 1$, the dual nature of the propagator enters the game. The function $K_1(x)\sim x^{-1/2}\exp [-x]$ introduces an exponential damping, that cuts off momenta larger than $1/L_0$.

The second intriguing property is that \eqref{eq:paddyprop} is an exact result both from a direct calculation in quantum field theory \cite{Pad97} and in string theory \cite{SSP03,SpF05,FSP06}. As we said in the previous section,  this is a piece of evidence for the model independent nature of Padmanabhan's path integral modification. In conclusion, both strings and the zero-point length capture a universal character of the Planck scale: the intrinsic nonlocality of the quantum spacetime.

\subsection{Some recent applications}

The most notable applications of Padmanabhan's propagator have been in the framework of electrodynamics \cite{GaN22} and general relativity \cite{NSW19,GJN22}.
A crucial starting point is the derivation of static potentials due to virtual particle exchange.
Such potentials are customary calculated via the generating functional of Green's function,
\begin{eqnarray}
U(r)&=&-\frac{1}{T}\ W[J],
\end{eqnarray}
with $T$ the time interval. 
For a standard massless field the potential has the usual $1/r$ profile. In case of \eqref{eq:paddyprop}, one finds  in the massless limit\footnote{The function $U(r)$ is actually a potential energy. It has to depend on the product of the charge (mass) of the source and of the probe. We omit such a product to keep the discussion valid for both electrostatic and gravitational potential energies. }
\begin{equation}
U(r)\propto \frac{1}{\sqrt{r^2 +L_0^2}}.
\end{equation}
The above potential is everywhere finite, in agreement with the concept of zero point length.
It is interesting to notice that the profile of the above potential is reflected in a non-local modification of the equation for Green's function, namely
\begin{equation}
{\nabla ^2} G\left( {{\bf x},{{\bf y} }} \right) =  - {L_0} \sqrt { - {\nabla ^2}}\, {K_1}\left( {{L_0}\sqrt { - 
{\nabla ^2}} } \right){\delta ^{\left( 3 \right)}}\left( {{\bf x} - {{\bf y}}} \right),   
\label{eq:greenfunction}
\end{equation}
where the Green's unmodified  function, $G_0 \left( {{\bf z},{{\bf z}^ \prime }} \right)$, reads
\begin{equation}
G_0 \left( {{\bf x},{{\bf y} }} \right)={\left[ {{L_0}\sqrt{- {\nabla ^2}}\,\,  {K_1}\left( {{L_0}\sqrt{ - {\nabla ^2}}  
} \right)} \right]^{ - 1}} G\left( {{\bf x},{{\bf y} }} \right). 
\end{equation}
The net effect of the zero point length $L_0$ is a smearing of the source in \eqref{eq:greenfunction}. The Dirac delta is replaced by a wider distribution, namely
\begin{eqnarray}
{\delta ^{\left( 3 \right)}}\left( {{\bf x} - {{\bf y}}} \right)&\longrightarrow &
\rho_\mathrm{eff}\left({{\bf x} - {{\bf y}}} \right) \nonumber\\ &\equiv & - {L_0} \sqrt { - {\nabla ^2}}\, {K_1}\left( {{L_0}\sqrt { - 
{\nabla ^2}} } \right){\delta ^{\left( 3 \right)}}\left( {{\bf x} - {{\bf y}}} \right) \nonumber\\
&\propto &  \frac{1}{\left(|{\bf x} - {{\bf y}}|^2+L_0^2\right)^{5/2}}.
\label{eq:effdensity}
\end{eqnarray}
To this purpose, non-local actions for electrodynamics has been derived in order to reproduce non-local Maxwell equations. The corresponding Lagrangian density reads \cite{GaN22}:
\begin{equation}
\mathcal{L}_\mathrm{ED} =
- \frac{1}{4}F_{\mu \nu }\, {\cal O} F^{\mu \nu }, \label{eq:NLMaxwell}
\end{equation}
where 
\begin{equation}
\mathcal{O} = {\left[ {{L_0}\sqrt \Delta\,\,  {K_1}\left( {{L_0}\sqrt \Delta  } \right)} \right]^{ - 1}}, \label{eq:NLoperator}
\end{equation}
is a non-local operator, with $\Delta\equiv {\partial _\mu }{\partial ^\mu }$. The profile of the non-local operator \eqref{eq:NLoperator} has been selected in order to fulfill two properties, i.e., it reproduces Green's function equation \eqref{eq:greenfunction}; it matches the Maxwell limit at low energy/large length scales $\mathcal{O}\approx 1$ for   $L_0\sqrt \Delta \ll 1.$ More importantly, it has been shown that the above Lagrangian is manifestly gauge invariant \cite{GaN22}.

Similarly to electrodynamics,  the modified equations for the gravity-matter system must display smeared source terms.  In the literature, there exists an array of actions leading to the non-local version of Einstein equations (see e.g. \cite{Bar03,HaW05,Mof11,MMN11,Mod12,BiM12}). Probably, the first proposal of such an action is due to Krasnikov \cite{Kra87} and Tomboulis \cite{Tom97} and reads:
\begin{eqnarray}
{\cal L}_\mathrm{G}& =& \sqrt{-g}\:\left\{\:\frac{\beta}{\kappa^2}R 
- \beta_2(R_{\mu\nu}R^{\mu\nu} - \frac{1}{3}R^2) + \beta_0R^2 + 
\tilde{\lambda} \right.\nonumber\\
 & &+\left.\left(R_{\mu\nu}\,h_2(-\frac{\Box}{\Lambda^2})
\,R^{\mu\nu} -\frac{1}{3}R\,h_2(-\frac{\Box}{\Lambda^2})\,R
\right) - R\,h_0(-\frac{\Box}{\Lambda^2})\,R
\:\right\}\nonumber\\
 & & -\frac{1}{2\xi}f^\mu[g]w(-\frac{\Delta}{\Lambda^2})
f_\mu[g] + \bar{c}^\mu M_{\mu\nu}c^\nu \label{eq:gact},
\end{eqnarray}
where $\Box=\nabla^\mu\nabla_\mu$ denotes the 
covariant  D'Alembertian,  
$f_\mu[g]$ is the gauge-fixing function with gauge-term weight 
$w$, $\bar{c}^\mu M_{\mu\nu}c^\nu$ is the Faddeev-Popov term, $\kappa^2= 16 \pi G$, $\Lambda$ is some energy scale, $\tilde{\lambda}$ is the cosmological constant and $h_0$, $h_2$ are non-polynomial entire functions. 
To implement the effects of the zero point length, one has to set the energy scale $\Lambda\sim 1/L_0^2$ and select the entire functions $h_0$, $h_2$. It is, however, not necessary to consider the full action. From truncated versions of \eqref{eq:gact}, one can derive the leading term of the non-local Einstein equation and identify the profile of $h_0$, $h_2$ at short scales  \cite{IMN13,Nic18}. This procedure is particularly convenient when a gauge field is coupled to gravity, as discussed in  \cite{GJN22}.


After solving the nonlocal Maxwell equations, the electric field can be displayed. In case of static conditions, it reads \cite{GaN22}:
\ba
\mathbf{E}(\mathbf{r})= -\frac{Q\ r}{4\pi \left(r^2+L_0^2\right)^{3/2}}\ \hat{\textbf{r}}.
\label{eq:elfield}
\ea 
The field is linearly vanishing at short distances and less intense of the corresponding field of the Maxwell theory. Nevertheless, the field equations admit a non-linear regime at short scales. From \eqref{eq:elfield} one finds that the Schwinger limit is violated for $r\sim L_0$, provided $1/L_0\geq 10$ GeV.

At larger distances, e.g., at the scale of atomic physics, the field equations are, however, linear. For this reason, the energy levels of the hydrogen atom have been used to set bounds on the parameter $L_0$ \cite{WoB19}. We can safely assume $1/L_0>10$ TeV, that is consistent with the non-observation of zero point length effects at  the LHC. In conclusion, after  taking into account both  bounds for $L_0$,  one expects  that any future observation of the zero point length in particle collisions should be accompanied by non-linear electrodynamics effects.

In standard general relativity, black holes are solutions of Einstein equation in (electro) vacuum. Conversely, in the presence of a zero point length, the non-locality of the action implies a non-vanishing source term. This is the tensor analogue of the density $\rho_\mathrm{eff}$ in \eqref{eq:effdensity} \cite{NSW19,GJN22}. For a static, spherically symmetric spacetime without charge, one can solve such non-local Einstein equations. The resulting  line element reads:
\begin{equation}
\label{eq:lineElem}
\der s^2
=- V(r)\, dt^2 +V^{-1}(r)\, dr^2 +r^2(d\theta^2+\sin^2\theta d\phi^2)
\end{equation}
with
\begin{equation}
V(r)
= 1 - \frac{2 M r^2}{\left(r^2+L_0^2\right)^{3/2}}.
\label{eq:bh}
\end{equation}
The above line element is curvature singularity free. Such a regularity is the gravitational analogue of the finiteness of the electrostatic field \eqref{eq:elfield}. The gravity case, however, discloses an additional important property. The region around the origin is a local de Sitter space. This represents an evidence for the presence of a regularized vacuum energy that violates conventional energy conditions.  The associated cosmological term provides an anti-gravity region that prevents the full collapse of the mass into a point-like distribution. The new phenomenology is uniquely governed by the parameter $L_0$ that effectively describes the fluctuations of the quantum manifold. A similar scenario has been obtained also in the context of noncommutative geometry \cite{NSS06,Nic09}, that is another byproduct of string theory. This fact further reinforces the universality of Padmanabhan's work.  

The Schwarzschild metric is an unstable configuration in the parameter space. As soon as charge or angular momentum are non-vanishing, the spacetime structure switches to a two-horizon, time-like singularity configuration. In a similar way, the introduction of a new parameter like $L_0$ implies the presence of two horizons, even in the absence of charge or angular momentum. This fact has terrific repercussions for the following reasons. 

When the outer and inner horizons coalesce, one has an extremal configuration. From a thermodynamic viewpoint, such a configuration represents an asymptotic zero temperature state. Accordingly, the end-point of the Hawking emission is no longer a runaway increase of the black hole temperature but a cold black hole remnant. The formation of such a remnant is preceded by a SCRAM, i.e.,  a positive heat capacity phase, during which the hole cools down\footnote{I introduced this term in 2009, by borrowing an acronym in use in nuclear power plant technology \cite{Nic09}. Originally the term SCRAM (``Safety control rod axe man'') was used by Fermi during the Pile-1 experiment in Chicago in 1942. It indicates the emergency shutdown of a nuclear reactor.}.
In other words, the presence of the zero point length makes the black hole thermodynamically stable. But there is something more. During the whole evaporation, the black hole \eqref{eq:bh} is colder of the corresponding Schwarzschild black hole with the same mass. In addition, the following inequality holds
\begin{equation}
T\ll \MP< M.
\end{equation}
This means that there is no relevant quantum back reaction. All the modification of the spacetime during the evaporation are already included in the new profile of the metric coefficients \eqref{eq:bh}.

This conclusion is of particular importance since it could represent an exhaustive explanation of the black hole life cycle. Indeed, both the Reissner-Nordström and Kerr black holes do admit a SCRAM phase similar to the black hole in \eqref{eq:lineElem}. They are, however, transient states. Via Hawking emission and Schwinger pair production the charge is shed and the rotation slows down. This means that both Reissner-Nordström and Kerr black holes decay into Schwarzschild black holes. Such black holes have well-known limitations for what concerns the geometry and the thermodynamics of the spacetime. Schwarzschild black holes can provide a reliable description only for masses larger than the Planck mass, $M\gg\MP$. On the other hand, the line element \eqref{eq:lineElem} undergoes a SCRAM and it is stable. This can be explained by the fact that, contrarily to the charge and the angular momentun,  the parameter, $L_0$, that guarantee the horizon extremisation, is not lost during the black hole evolution. For this reason, \eqref{eq:lineElem} represents one of the few reliable ultraviolet completions of the Schwarzschild metric.

\section{Conclusions}
\label{sec:concl}

Quantum gravity can be considered the final step along the path towards a full understanding of the Universe and its matter content. After an overview about the open problems connected to quantum gravity, we focused on the inspirational work of Padmanabhan about the duality of the path integral. Such a duality is strictly connected to the presence of a zero point length $L_0$ over a quantum manifold \cite{Pad97}. We have stressed the good properties of such a result, namely its universality,  model independence and connection with a candidate theory of everything like string theory \cite{SSP03,SpF05,FSP06}.

In the second part of this paper, we have exploited an additional good feature of Padmanabhan's propagator: its predictive power. Indeed, in contrast to many approaches to quantum gravity, calculations in the presence of a zero point length can be carried on, in most cases, just by using analytic methods. To this purpose, we have presented two recent results: the derivation of a theory of electromagnetism  and a static, neutral black hole spacetime. In both cases, the main goal of the path integral duality program has been achieved: The zero point length efficiently works as a natural ultraviolet cutoff of geometric origin. Short scale divergence of the electrostatic  field and curvature singularities of black hole spacetimes are removed.

The most intriguing consequences descending from the zero point length are, however, on the side of black hole thermodynamics. Even in the case of neutral, spinless black holes, the spacetime admits two horizons. This implies the existence of a positive heat capacity, cooling down phase towards a zero temperature remnant configuration. In practice all the limitations of the semiclassical description of the black hole life cycle are overcome.  We stress that such a scenario is a consequence of the presence of just a single parameter $L_0$.

Further repercussions of the path integral duality are currently the subject of investigations in the context of charged and charged, spinning black holes \cite{GJN22}. Apart from the manifold regularity, the  zero point length may help to shed light about the duration of the phases preceding the neutral, spherically symmetric one. There are hopes to have robust indications about the balding and spin-down phases, their duration and superposition. 

The work of Padmanabhan will certainly affect the path towards a successful understanding of subtle aspects of quantum gravity in the close and distant future.

\section{My memories of Paddy}
\label{sec:IandPaddy}

I first met Padmadabhan in 2009 at the University of Geneva. At that time, we were both visiting  Ruth Durrer and her group. Even before our meeting, Paddy, as everybody called him, was already known to me. He was a very well known scientist, who has already conjectured some original paradigms and ideas.  Our acquaintance, however, revealed interesting new aspects of his  personality. I immediately noticed his keen interest and curiosity about everything. He also had a never ending energy in writing papers and making the point during  brainstorming sessions.
He was also very helpful and willing to support the work of the younger generations with punctual comments and references. 

Four years later, I organized in Frankfurt a conference named after Karl Schwarzschild. Padmanabhan was at the top of my list
of candidates for a  keynote slot. I knew, however, that the visit in Frankfurt would have costed him some bothers due to the heavy paperwork for visa related matters. For this reason, I was happily surprised and pleased when he accepted my invitation.  During the conference,  his participation was an additional evidence of dedication and commitment to science. His keynote titled ``Emergent gravity and the cosmological constant'' was extremely well received \cite{Pad16}. The structure of his presentation was actually impeccable. At the very beginning, there was a pedagogical overview of the main issues connected to gravity. In the second part of the talk, Padmanabhan focused on recent developments about gravity as an emergent phenomenon. In such a way, he was able to engage both the non-expert audience and scientists working in the field.

Interestingly, one of the key points of Padmanabhan's talk in Frankfurt is connected to the main subject of the current paper, namely, the  necessity for a quantum theory of gravity. The words of one his slides are reported at the beginning of this paper.

After the inaugural Karl Schwarzschild Meeting, I and Padmanabhan have regularly been in touch, mostly via Email. In 2015, he and his wife Vasanthi   visited me in Frankfurt, on the way from Pune to  Zürich, where their daughter Hamsa was working as a PhD student. In that occasion, he held a talk titled 
``Gravity and the Cosmos'' for the colloquium series at the Frankfurt Institute for Advanced Studies.
More recently, we had an intensive Email exchange about my work on the applications of the path integral duality, a property he conjectured in 1997. 

In conclusion, Padmanabhan has left an indelible mark in the history of classical and quantum gravity.  His guidance and lessons will not be lost.

{\small
\section*{Acknowledgments}

The work of P.N. has partially been supported by GNFM, Italy's National Group for Mathematical Physics. P.N. is grateful to Sumanta Chakraborty,  Dawood Kothawala, Sudipta Sarkar, Amitabh Virmani for the invitation to submit a contribution to the
Topical Collection (TC) ``In Memory of Prof. T. Padmanabhan'' of the journal General Relativity and Gravitation (GERG). P.N. is grateful to Athanasios Tzikas for the support in drawing the picture.

\subsubsection*{Data Availability Statement}

Data sharing not applicable to this article as no datasets were generated or analysed during the current study.
}

\end{document}